\documentclass[journal, twoside]{IEEEtran}
\usepackage{cite}
\usepackage{graphicx}
\usepackage{footnote}
\usepackage{bbding}
\usepackage{pifont}
\usepackage{multirow}
\usepackage[table,xcdraw]{xcolor}
\usepackage{diagbox}
\usepackage[flushleft]{threeparttable}
\makesavenoteenv{tabular}
\usepackage{tabu}
\usepackage{longtable}
\usepackage{cite}
\usepackage{epstopdf}
\usepackage{color}
\usepackage{enumitem}
\usepackage{booktabs}
\usepackage{subfigure}
\usepackage{amsmath}

\setlist{parsep=0pt,listparindent=\parindent}

\graphicspath{ {image/} }

\ifCLASSINFOpdf
\else
\fi

\begin{document}
%
\title{E-LSTM-D: A Deep Learning Framework for Dynamic Network Link Prediction}

\author{Jinyin~Chen,
        Jian~Zhang,
        Xuanheng~Xu,
        Chengbo~Fu,
        Dan~Zhang,
        Qingpeng~Zhang~\IEEEmembership{IEEE Member},
        and~Qi~Xuan~\IEEEmembership{IEEE Member}
\thanks{This work is partially supported by Zhejiang Natural Science Foundation (LY19F020025, LR19F030001), National Natural Science Foundation of China (61502423, 61572439, 11505153), Zhejiang University Open Fund (2018KFJJ07), Zhejiang Science and Technology Plan Project (LGF18F030009, 2017C33149), and State Key Laboratory of Precision Measuring Technology and Instruments Open Fund. (\emph{Corresponding author: Qi Xuan.})}
\thanks{J. Chen, J. Zhang, X. Xu, C. Fu, D. Zhang, and Q. Xuan are with the Zhejiang University of Technology, Hangzhou 310023, China (e-mail: chenjinyin@zjut.edu.cn; zj\_1994@outlook.com; 2111603112@zjut.edu.cn; cbfu@zjut.edu.cn; danzhang@zjut.edu.cn; xuanqi@zjut.edu.cn).}
\thanks{Q. Zhang is with the City University of Hong Kong, Hong Kong, China (e-mail:  qingpeng.zhang@cityu.edu.hk).}}


\maketitle

\begin{abstract}
Predicting the potential relations between nodes in networks, known as link prediction, has long been a challenge in network science. However, most studies just focused on link prediction of static network, while real-world networks always evolve over time with the occurrence and vanishing of nodes and links. Dynamic network link prediction thus has been attracting more and more attention since it can better capture the evolution nature of networks, but still most algorithms fail to achieve satisfied prediction accuracy. Motivated by the excellent performance of Long Short-Term Memory (LSTM) in processing time series, in this paper, we propose a novel Encoder-LSTM-Decoder (E-LSTM-D) deep learning model to predict dynamic links end to end. It could handle long term prediction problems, and suits the networks of different scales with fine-tuned structure. To the best of our knowledge, it is the first time that LSTM, together with an encoder-decoder architecture, is applied to link prediction in dynamic networks. This new model is able to automatically learn structural and temporal features in a unified framework, which can predict the links that never appear in the network before. The extensive experiments show that our E-LSTM-D model significantly outperforms newly proposed dynamic network link prediction methods and obtain the state-of-the-art results.
\end{abstract}


\IEEEpeerreviewmaketitle

\section{Introduction}
\IEEEPARstart{N}{etworks} are often used to describe complex systems in various areas, such as social sicence~\cite{ediger2010massive,fu2017pinning}, biology~\cite{wang2017investigating}, electric system~\cite{GAO2012391} and economics\cite{kazemilari2015correlation} etc. And the vast majority of the real world systems evolve with time, which can be modeled as dynamic networks~\cite{sun2017complex, liu2018optimizing}, where the nodes may come and go and the links may vanish and recover as time goes by. Links, representing the interactions between different entities, are of particular significance in the analysis of dynamic networks.


Link prediction of a dynamic network~\cite{ibrahim2015link,xuan2015temporal} tries to predict the future structure of the network based on the historical data, which helps us better understand network evolution and further the relationships between topologies and functions. For instance, in online social networks~\cite{xuan2018social,xuan2018modern,fu2018link}, we can predict which links are going to be established in the near future. It means that we can infer with what kind of people, or even which particular one, the target user probably makes friends base on their historical behaviors. It can also be applied to the studies on disease contagions~\cite{lentz2016disease}, protein-protein interactions~\cite{theocharidis2009network} and many other fields where the evolution matters.


Similarity indices, like Common Neighbor~(CN)~\cite{newman2001clustering} and Resource Allocation Index~(RA)~\cite{zhou2009predicting}, are widely used in link prediction of static networks~\cite{lu2011link}, but they can hardly deal with the changes of the network structure directly. To learn temporal dependencies, Yao et al.~\cite{yao2016link} assigned time-varied weights to previous graphs and then execute link prediction task using the refined CN which considers the neighbors within two hops. Similarly, Zhang et al.~\cite{zhang2017efficient} proposed an improved RA based dynamic network link prediction algorithm, which updates the similarity between pairwise nodes when the network structure changes. These methods, however, mostly depend on simple statistics of networks and thus cannot effectively deal with high non-linearity.

In order to tackle this problem, a bunch of network embedding techniques were proposed to learn the representations of networks that can preserve high-order proximity. Random walk based method, such as DeepWalk~\cite{perozzi2014deepwalk} and node2vec\cite{grover2016node2vec}, sample sequences of nodes and get node vectors by applying skip-gram. Furthermore, with the development of deep learning~\cite{han2018deep, xuan2018automatic,xuan2018evolving}, methods like structural deep network embedding~(SDNE)~\cite{wang2016structural} and Graph Convolution Network~(GCN)~\cite{kipf2016semi}, can automatically learn node representations end to end. The embedding vectors ensure the nodes of similar structural properties stay close in the embedding space. These embedding methods are powerful but still lack the ability of analyzing the evolution of networks. To learn such temproal dependencies, some recent works take the evolution of network into consideration. Ahmed et al.~\cite{ahmed2016sampling} assigned damping weights to each snapshots, ensuring that more recent snapshots are more important, and combine them into a weighted graph to do local random walk. As an extension of~\cite{ahmed2016sampling}, Ahmed and Chen~\cite{ahmed2016efficient} proposed Time Series Random Walk (TS-RW) to integrate temporal and global information. There are also some methods based on Restrict Boltzmann Machine (RBM), which regard the evolution of network as a special case of Markov random field with two-layer variables. Conditional temporal RBM~\cite{li2014deep}, namely ctRBM, considers not only neighboring connections but also temporal connections, and thus has the ability to predict future links. Zhou et al.~\cite{zhou2018dynamic} modeled the network evolution as a triadic closure process, which however is limited to undirected networks. Following the idea of SDNE, Li et al.~\cite{li2018deep} used Gated Recurrent Unit (GRU)~\cite{chung2014empirical} as encoder to learn both spatial and temporal information. Most of these combinations, however, are limited to predicting the added links, which only reflects a part of network evolution. Moreover, they have to obtain a representation of links and then train a binary classification model, which is less unified.

In this paper, we address the problem of predicting the global structure of networks in the near future, focusing on the links that are going to appear or disappear. We propose a novel end-to-end Encoder-LSTM-Decoder (E-LSTM-D) deep learning model for link prediction in dynamic networks, which takes the advantages of encoder-decoder architecture and a stacked Long Short-Term Memory (LSTM). The model thus can effectively handle the problems of high dimension, non-linearity and sparsity. Due to the encoder-decoder architecture, the model can automatically learn representations of networks, as well as reconstruct a graph on the grounds of the extracted information. Relatively low dimensional representations for the sequences of graphs can be well learned from the stacked LTSM module placed right behind the encoder. Considering that network sparsity may seriously affect the performance of the model, we amplify the effect of existing links at the training process, enforcing the model to account for the existing links more than missing/nonexistent ones. We conduct comprehensive experiments on five real-world datasets. The results show that our model significantly outperforms the current state-of-the-art methods. In particular, we make the following main contributions.
\begin{itemize}
\item We propose a general end-to-end deep learning framework, namely E-LSTM-D, for link prediction in dynamic networks, where the encoder-decoder architecture automatically learns representations of networks and the stacked LSTM module enhances the ability of learning temporal features.
\item Our newly proposed E-LSTM-D model is competent to make long term prediction tasks with only slight drop of performances; It suits the networks of different scales by fine tuning the model structure, i.e., changing the number of units in different layers; Besides, it can predict the links that are going to appear or disappear, while most existing methods only focus on the former.
\item We define a new metric, \emph{Error Rate}, to measure the performance of dynamic network link prediction, which is a good addition to the Area Under the ROC Curve (AUC), so that the evaluation is more comprehensive.
\item We conduct extensive experiments, comparing our E-LSTM-D model with five baseline methods on various metrics. It is shown that our model outperforms the others and obtain the state-of-the-art results.
\end{itemize}

The rest of paper is organized as follows. In Section~\ref{Methodology}, we provide a rigorous definition of dynamic network link prediction and a detailed description of our E-LSTM-D model. Comprehensive experiments are presented in Section~\ref{Exp}, with the results carefully discussed. Finally, we conclude the paper and outline some future works in Section~\ref{Con}.

\begin{figure}[!t]
\centering
\includegraphics[width=1.\linewidth]{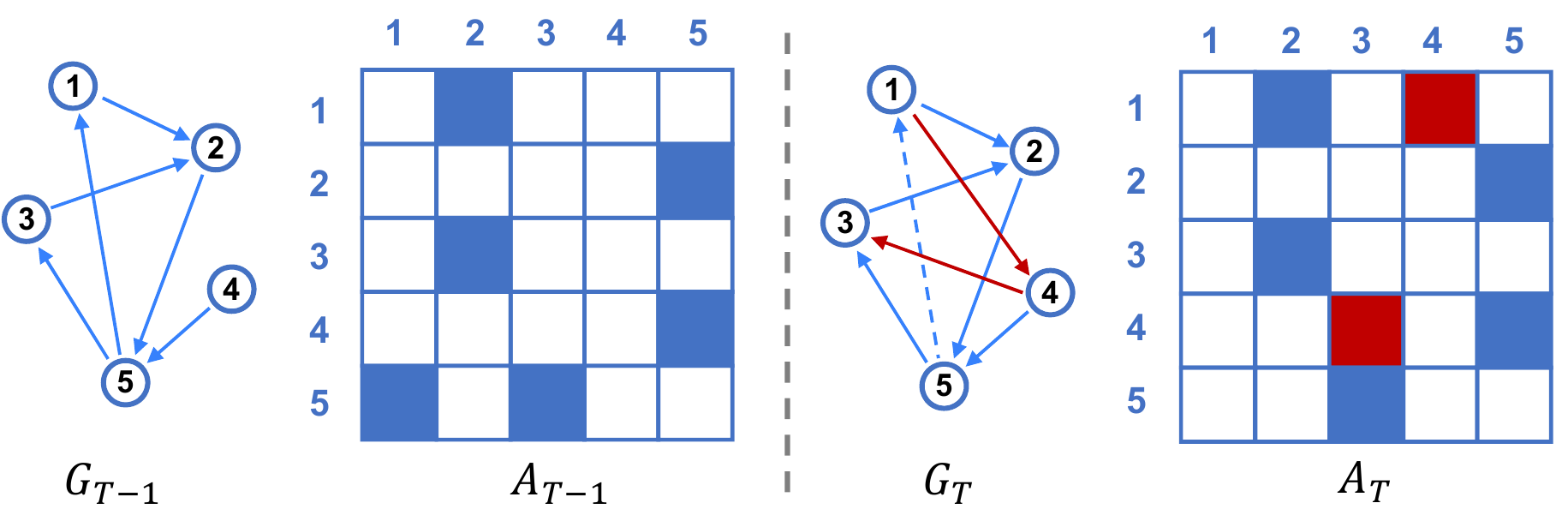}
\caption{\textbf{An illustration of network evolution.}The structure of the network changes overtime. At time $T$, ${E}_{(1,4)}$ and ${E}_{(4,3)}$ emerge while ${E}_{(5,1)}$ vanishes, which is reflected in the change of $A$, with those elements equal to 1 represented by filled squares.}
\label{fig:adjacency}
\end{figure}

\section{Methodology\label{Methodology}}
In this section, we will introduce our E-LSTM-D model used to predict the evolution of dynamic networks.

\subsection{Problem Definition}
A dynamic network is modeled as a sequence of snapshot graphs taken at a fixed interval.

\newtheorem{Dynamic network}{\textbf{\textsc{Definition}}}
\newtheorem{Link prediction}[Dynamic network]{\textbf{\textsc{Definition}}}

\begin{Dynamic network}[Dynamic Networks]
Given a sequence of graphs, \{${G}_1$, ..., ${G}_T$\}, where ${G}_k=(V, {E}_{k})$ denotes the $k^{th}$ snapshot of a dynamic network. Let $V$ be the set of all vertices and ${E}_{k}\subseteq{{V}\times{V}}$ the temporal links within the fixed timespan $[{t}_{k-1}, {t}_{k}]$. The adjacency matrix of ${G}_{k}$ is denoted by ${A}_{k}$ with the element ${a}_{k;i,j}=1$ if there is a directed link from $v_i$ to $v_j$ and ${a}_{k;i,j}=0$ otherwise.
\end{Dynamic network}

In a static network, link prediction aims to find edges that actually exist according to the distribution of observed edges. Similarly, link prediction in a dynamic network makes full use of the information extracted from previous graphs to reveal the underlying network evolving patterns, so as to predict the future status of the network. Since the adjacency matrix can precisely describe the structure of a network, it is ideal to use it as the input and output of the prediction model. We could infer ${G}_{t}$ just based on ${G}_{t-1}$, due to the strong relationship between the successive snapshots of the dynamic network. However, the information contained in ${G}_{t}$ may be too little to do precise inference. In fact, not only the structure itself but also the structure change overtime matters in the network evolution. Thus, we prefer to use a sequence of length $N$, i.e., \{${G}_{t-N}$, ...,${G}_{t-1}$\}, to predict ${G}_{t}$.

\begin{Link prediction}[Dynamic Network Link Prediction]
Given a sequence of graphs with length $N$, $S$=\{${G}_{t-N}$, ..., ${G}_{t-1}$\}, Dynamic Network Link Prediction (DNLP) aims to learn a function that maps the input sequence $S$ to ${G}_{t}$.
\end{Link prediction}

The structure of a dynamic network evolves with time. As shown in Fig.~\ref{fig:adjacency}, some links may emerge while some others may vanish, which can be reflected by the changes of the adjacency matrix overtime. The goal is to find the links of the network that are most likely to appear or disappear at the next timespan. Mathematically, it can also be interpreted as an optimization problem of finding a matrix, whose element is either 0 or 1, that can best fit the ground truth.

\begin{figure*}[!t]
\centering
\includegraphics[width=1.\linewidth]{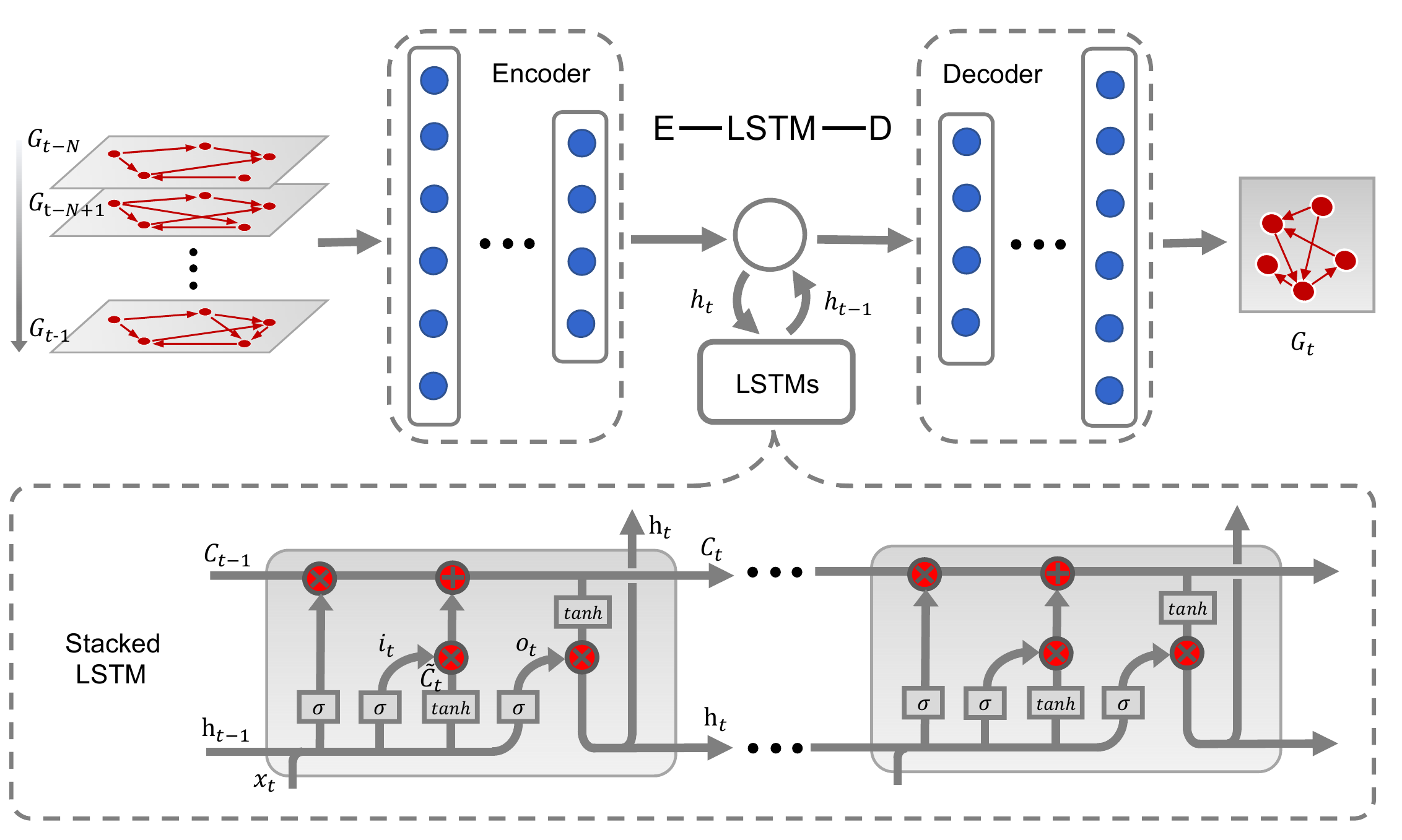}
\caption{\textbf{The overall framework of E-LSTM-D model}. Given a sequence of graphs with length $N$, \{$G_{t-N}$, $G_{t-N+1}$, $\cdots$, $G_{t-1}$\}, the encoder maps them into a lower dimensional latent space. Each graph is transformed into a matrix that represents the structural features. And then the stacked LSTM, composed of multiple LSTM cells, learns network evolution patterns from the extracted features. The decoder projects the received feature maps back to the original space to get $G_{t}$. Here, $\sigma$ in LSTM cells is an activation function and we use sigmoid in this paper.}
\label{fig:framework}
\end{figure*}

\subsection{E-LSTM-D Framework}
Here, we propose a novel deep learning model, namely E-LSTM-D, combining the architecture of encoder-decoder and stacked LSTM, with the overall framework shown in Fig.~\ref{fig:framework}. Specifically, the encoder is placed at the entrance of the model to learn the highly non-linear network structures and the decoder converts the extracted features back to the original space. Such encoder-decoder architecture is capable of dealing with spatial non-linearity and sparsity, while the stacked LSTM between the encoder and decoder can learn temporal dependencies. The well designed end-to-end model thus can learn both structural and temporal features and do link prediction in a unified way.

We first introduce terms and notations that will be frequent used later, all of which are listed in TABLE~\ref{table:notation}. Other notations will be explained along with the corresponding equations. Notice that a single LSTM cell can be regarded as a layer, in which the terms with subscript \textit{f} are the parameters of forget gate, the terms with subscripts \textit{i} and \textit{C} are the parameters of input gate, and those with subscript \textit{o} are the parameters of output gate.

\subsubsection{Encoder-decoder architecture}
Autoencoder can efficiently learn representations of data in an unsupervised way. Inspired by this, we place an encoder at the entrance of the model to capture the highly non-linear network structure and a graph reconstructor at the end to transform the latent features back into a matrix of fixed shape. Here, however, the whole process is supervised, which is different from autoencoder, since we have labeled data (${A}_{t})$ to guide the decoder to build matrices that can better fit the target distributions. In particular, the encoder, composed of multiple non-linear perceptions, projects the high dimensional graph data into a relatively lower dimensional vector space. Therefore, the obtained vectors could characterize the local structure of vertices in the network. This process can be characterized as
\begin{equation}
\label{eq:encoder}
\begin{split}
& y^{(1)}_{e;i} = \mathit{ReLU}(W^{(1)}_{e}s_{i} + b^{(1)}_{e}) \\
& y^{(k)}_{e;i} = \mathit{ReLU}(W^{(k)}_{e}y^{(k-1)}_{e;i} + b^{(k)}_{e}) \\
& Y^{(k)}_{e} = [y^{(k)}_{e;0}, \cdots, y^{(k)}_{e;N-1}],  \\
\end{split}
\end{equation}
where $s_{i}$ represents $i^{th}$ graph in the input sequence $S$. For an input sequence, each encoder layer processes every term separately and then concatenates all the activations in the order of time. Here, we use $\mathit{ReLU}$ as the activation function for each encoder/decoder layer to accelerate convergence.

\begin{table}[!t]
\renewcommand{\arraystretch}{1.6}
\caption{Terms and notations used in the framework.}
\label{table:notation}
\centering
\begin{tabular}{lr}
\hline
Symbol & Definition \\
\hline \hline
$K$                                & number of encoder/decoder layers\\
$L$                                & number of LSTM cells\\
$\hat{A}_t$                        & output of the decoder\\
$H$                                & output of the stacked LSTM\\
${Y}^{(k)}_{e}$, ${Y}^{(k)}_{d}$   & output of ${k}^{th}$ encode/decoder layer\\
${W}^{(k)}_{e}$, ${W}^{(k)}_{d}$   & weight of ${k}^{th}$ encode/decoder layer\\
${b}^{(k)}_{e}$, ${b}^{(k)}_{d}$   & bias of ${k}^{th}$ encoder/decoder layer\\
${W}^{(l)}_{f, i, C, o}$           & weight of ${l}^{th}$ LSTM layer\\
${b}^{(l)}_{f, i, C, o}$           & bias of ${l}^{th}$ LSTM layer\\
\hline
\end{tabular}
\end{table}

The decoder with the mirror structure of the encoder receives the latent features and maps them into the reconstruction space under the supervision of $A_{t}$, represented by
\begin{equation}
\label{eq:decoder}
\begin{split}
& Y^{(1)}_{d} = \mathit{ReLU}(W^{(1)}_{d}H + b^{(1)}_{d}) \\
& Y^{(k)}_{d} = \mathit{ReLU}(W^{(k)}_{d}Y^{(k-1)}_{d} + b^{(k)}_{d}), \\
\end{split}
\end{equation}
where $H$ is generated by the stacked LSTM and represents the features of the target snapshot rather than a sequence of features of all previous snapshots used in the encoder. Another difference is the last layer of the decoder, or the output layer, uses sigmoid as the activation function rather than $\mathit{ReLU}$. And the number of units of the output layer always equals to the number of nodes.

\subsubsection{Stacked LSTM}
Although encoder-decoder architecture could deal with the high non-linearity, it is not able to capture the time-varying characteristics. LSTM~\cite{gers1999learning}, as a special kind of recurrent neural network~(RNN)~\cite{lipton2015critical, han2019seqviews2seqlabels}, can learn long-term dependencies and is introduced here to solve this problem. An LSTM consists of three gates, i.e., a forget gate, an input gate and an output gate. The first step is to decide what information is going to be thrown away from previous cell state. The operation is performed by the forget gate, which is defined as
\begin{equation}
f_t = \sigma(W_f\cdot[h_{t-1}, {Y}^{(K)}_{e}] + b_f),
\end{equation}
where $h_{t-1}$ represents the output at time $t-1$. Then the input gate decides what new information should be added to the cell state. First, a sigmoid layer decides what information the input contains, $i_t$, should be updated. Second, a tanh layer generates a vector of candidate state values, $\tilde{C}_{t}$, which could be added to the cell state. The combination of $i_t$ and $\tilde{C}_{t}$ represents the current memory that can be used for updating $C_t$. The operation is defined as
\begin{equation}
\begin{split}
& i_t = \sigma(W_i\cdot[h_{t-1}, {Y}^{(K)}_{e}] + b_i) \\
& \tilde{C}_{t} = tanh(W_C\cdot[h_{t-1}, {Y}^{(K)}_{e}] + b_C) \\
& C_t = f_t\ast{C_{t-1}} + i_t\ast{\tilde{C}_t}. \\
\end{split}
\end{equation}
Taking the benefit of the forget gate and the input gate, LSTM cell can not only store long-term memory but also filter out the useless information. The output of LSTM cell is based on $C_t$ and it is controlled by the output gate which decides what information, $o_t$, should be exported. The process is described as
\begin{equation}
\begin{split}
& o_t = \sigma(W_o\cdot[h_{t-1}, {Y}^{(K)}_{e}] + b_o) \\
& h_t = o_t\ast{tanh(C_t)}. \\
\end{split}
\end{equation}
A single LSTM cell is capable of learning time dependencies, but a chain-like LSTM module, namely stacked LSTM, is more suitable for processing time sequence data. Stacked LSTM consists of multiple LSTM cells that take signals as input in the order of time. We place the stacked LSTM between the encoder and the decoder to learn the patterns under which the network evolves. After receiving the features extracted at time $t$, the LSTM module turns them into ${h}_{t}$ and then feed ${h}_{t}$ back to the model at next training step. It helps the model make use of the remaining information of previous training data. It should be always noticed that the numbers of units in encoder, LSTM cells and decoder vary when $N$ changes. The larger $N$, the more units we need in the model.

The encoder at the entrance could reduce the dimension for each graph and thus keep the computation of the stacked LSTM at a reasonable cost. And the stacked LSTM which is advanced at dealing with temporal and sequential data is supplementary to the encoder in turn.

\subsection{Balanced Training Process}
${L}_2$ distance, often applied in regression, can measure the similarity between two samples. But if we simply use it as loss function in the proposed model, the cost could probably not converge to an expected range or result in overfitting due to the sparsity of the network. There are far more zero elements than non-zero elements in $A_t$, making the decoder appeal to reconstruct zero elements. To address this sparsity problem, we should focus more on those existing links rather than nonexistent links in back propagation. We define a new loss function as
\begin{equation}
\label{eq:loss_l2}
\begin{split}
\mathcal{L} &= \sum_{i=1}^{n} \sum_{j=1}^{n} (a_{t;i,j} - \hat{a}_{t;i,j})*p_{i,j}\\
            &= {\parallel (A_{t} - \hat{A}_{t}) \odot{P} \parallel}_{F}^2,
\end{split}
\end{equation}
where $\odot$ means the Hadamard product. For each training process, $p_{i,j}=1$ if $a_{t;i,j}=0$ and $p_{i,j}=\beta>1$ otherwise. Such penalty matrix exerts more penalty on non-zero elements so that the model could avoid overfitting to a certain extent. And we finally use the mixed loss function
\begin{equation}
\label{eq:loss_total}
\mathcal{L}_{total} = \mathcal{L} + \alpha\mathcal{L}_{reg},
\end{equation}
where $\mathcal{L}_{reg}$, defined in Eq.~(\ref{eq:loss_reg}), is a $\mathcal{L}_{2}$ regularizer to prevent the model from overfitting and $\alpha$ is a tradeoff parameter.
\begin{equation}
\label{eq:loss_reg}
\begin{split}
\mathcal{L}_{reg} = &\frac{1}{2} \sum_{k=1}^{K} \bigg({\| W^{(k)}_{e} \|}_{F}^2 + {\| \hat{W}^{(k)}_{d} \|}_{F}^2 \bigg)\\
                    & + \frac{1}{2} \sum_{l=1}^{L} \bigg({\|W^{(l)}_{f}\|}_{F}^2 + {\|W^{(l)}_{i}\|}_{F}^2\\
                    & +  {\| W^{(l)}_{C} \|}_{F}^2 +{\| W^{(l)}_{o} \|}_{F}^2 \bigg).
\end{split}
\end{equation}

The value of each element in $A$ is either 0 or 1. The output data, however, are not one-hot encoded. They are decimals and could go to infinity or move towards the opposite direction theoretically. In order to get a valid adjacency matrix, we impose a sigmoid function at the output layer and then modify the values to 0 and 1 with 0.5 as the demarcation point. That is, there exists a link between $i$ and $j$ if $\hat{a}_{t;i,j}\geq0.5$ and there is no link otherwise. To optimize the proposed model, we should first make a forward propagation to obtain the loss and then do back propagation to update all the parameters. In particular, the key operation is to calculate the partial derivative of
$\partial{\mathcal{L}_{total}}/\partial{W^{(k)}_{e,d}}$ and $\partial{\mathcal{L}_{total}}/\partial{W^{(l)}_{f, i, C, o}}$.

We would like to take the calculation of ${\partial{\mathcal{L}_{total}}}/{\partial{W^{(k)}_{e}}}$ for instance. Taking partial derivative with respect to $W^{(k)}_{e}$ of Eq.~(\ref{eq:loss_total}), we have
\begin{equation}
\begin{split}
\frac {\partial{\mathcal{L}_{total}}}{\partial{W^{(k)}_{e}}} &= \frac {\partial{\mathcal{L}}}{\partial{W^{(k)}_{e}}} + \alpha{\frac {\partial{\mathcal{L}_{reg}}}{\partial{W^{(k)}_{e}}}}\\
                                                         &= \frac {\partial{\mathcal{L}}}{\partial{A_t}}\cdot{\frac {\partial{A_t}}{\partial{W^{(k)}_{e}}}} + \alpha{\parallel{W^{(k)}_{e}}\parallel}_{F}.
\end{split}
\end{equation}
According to Eq.~(\ref{eq:loss_l2}), we can easily obtain
\begin{equation}
\frac {\partial{\mathcal{L}}}{\partial{A_t}} = 2(A_t-\hat{A}_t)\odot{P}.
\end{equation}
To calculate ${\partial{A_t}}/{\partial{W^{(k)}_{e}}}$, we should iteratively take partial derivative with respect to $\partial{W^{(k)}_{e}}$ on both sides of Eq.~(\ref{eq:encoder}). After getting $\partial{\mathcal{L}_{total}}/\partial{W^{(k)}_{e}}$, we update the weight by

\begin{equation}
W^{(k)}_{e} = W^{(k)}_{e}-\lambda{\frac {\partial{\mathcal{L}_{total}}}{\partial{W^{(k)}_{e}}}},
\end{equation}
where $\lambda$ is the learning rate which is set as 1e-3 in the following experiments.

As for ${\partial{A_t}}/{\partial{W^{(k)}_{d}}}$ and $\partial{\mathcal{L}_{total}}/\partial{W^{(l)}_{f, i, C, o}}$, the calculation of partial derivative almost follows the same procedure, though it is a little more complicated when it comes to the weights in LSTM cells. This is because the recurrent network makes use of cell states at every forward propagation cycle.

\section{Experiments\label{Exp}}
The proposed E-LSTM-D then is evaluated on five benchmark datasets, compared with four baseline methods.

\subsection{Datasets}
We perform the experiments on five real-world dynamic networks, all of which are human contact networks, where nodes denote humans and links stand for their contacts. The contacts could be face-to-face proximity, emailing and so on. The detailed descriptions of these datasets are listed below.
\begin{itemize}
\item \textbf{\textsc{contact}}~\cite{konect:2017:contact}: It is a human contact dynamic network of face-to-face proximity. The data are collected through the wireless devices carried by people. A link between person $s$ (source) and $t$ (target) emerges along with a timestamp if $s$ gets in touch with $t$. The data are recorded every 20 seconds and multiple edges may be shown at the same time if multiple contacts are observed in a given interval.
\item \textbf{\textsc{enron}}~\cite{nr} and \textbf{\textsc{radoslaw}}~\cite{konect:radoslaw}: They are email networks and each node represents an employee in a mid-sized company. A link occurs every time an e-mail sent from one to another. \textsc{enron} records email interactions for nearly 6 months and \textsc{radoslaw} lasts for nearly 9 months.
\item \textbf{\textsc{fb-forum}}~\cite{konect:2017:facebook-wosn-wall}: The data were attained from a Facebook-like online forum of students at University of California, Irvine, in 2004. It is an online social network where nodes are users and links represent interactions (e.g., messages) between them. The records span more than 5 months.
\item \textbf{\textsc{lkml}}~\cite{konect:2017:lkml-reply}: The data were collected from linux kernel mailing list. The nodes represent users which are identified by their email addresses and each link donates a reply from one user to another. We only focus on the 2210 users that were recorded from 2007-01-01 to 2007-04-01 and then construct a dynamic network based on the links between these users that appeared from 2007-04-01 to 2013-12-01.
\end{itemize}
All the experiments are implemented in both long-term and short-term networks. The basic statistics of the five datasets are summarized in TABLE~\ref{dataset}.

\begin{table}[!t]
\newcommand{\tabincell}[2]{\begin{tabular}{@{}#1@{}}#2\end{tabular}}
\renewcommand{\arraystretch}{1.3}
\caption{The basic statistics of the five datasets.}
\label{dataset}
\centering
\begin{tabular}{cccccc}
\hline \hline
Dataset                 & $|V|$  & $|{E}_{T}|$  & $\bar{d}$ & $d_{max}$ & \tabincell{l}{Timespan\\~~~(days)}\\
\hline
\textsc{contact}        & 274    & 28.2K        & 206.2     & 2,092      & 4.0\\
\textsc{enron}          & 151    & 50.5K        & 669.8     & 1,841      & 164.5\\
\textsc{radoslaw}       & 167    & 82.9K        & 993.1     & 9,053      & 271.2\\
\textsc{fb-forum}       & 899    & 50.5K        & 669.8     & 5,177      & 164.5\\
\textsc{lkml}           & 2210   & 422.4K       & 34.6      & 47,995     & 2,436.3 \\

\hline \hline
\end{tabular}
\end{table}

Before training, we take snapshots for each dataset at a fixed interval and then sort them in an ascending order of time. Considering that the connections between people are probably temporary, we remove the links that do not show up again in the following 8 intervals and the length of each interval may vary for different timespan. To obtain enough samples, we split each dataset into 320 snapshots with different intervals and set $N=10$. In this case, $\{G_{t-10},\ldots,G_{t-1},G_t\}$ is treated as a sample with the first ten snapshots as the input and the last one as the output. As a result, we can get 310 samples in total. We then group the first 230 samples, with $t$ varying from 11 to 240, as the training set, and the rest 80 samples, with $t$ varying from 241 to 320, as the test set.

\subsection{Baseline Methods}
To validate the effectiveness of our E-LSTM-D model, we compare it with node2vec, as a widely used baseline network embedding method, as well as four state-of-the-art DNLP methods that could handle time dependencies, including Temporal Network Embedding (TNE)~\cite{zhu2016scalable}, conditional temporal RBM (ctRBM)~\cite{li2014deep}, Gradient boosting decision tree based Temporal RBM (GTRBM)~\cite{li2018restricted} and Deep Dynamic Network Embedding (DDNE)~\cite{li2018deep}. In particular, the five baselines are introduced as follows.
\begin{itemize}
\item \textbf{node2vec}~\cite{grover2016node2vec}: As a network embedding method, it maps the nodes of a network from a high dimensional space to a lower dimensional vector space. A pair of nodes tend to be connected with a higher probability, i.e., they are more similar, if the corresponding vectors are of shorter distance.
\item \textbf{TNE}~\cite{zhu2016scalable}: It models network evolution as a Markov process and then use the matrix factorization to get the embedding vector for each node.
\item \textbf{ctRBM}~\cite{li2014deep}: It is a generative model based on temporal RBM. It first generates a vector for each node based on temporal connections and predict future linkages by integrating neighbor information.
\item \textbf{GTRBM}~\cite{li2018restricted}: It takes the advantages of both tRBM and GBDT to effectively learn the hidden dynamic patterns.
\item \textbf{DDNE}~\cite{li2018deep}: Similar to autoencoder, it uses a GRU as an encoder to read historical information and decodes the concatenated embeddings of previous snapshot into future network structure.
\end{itemize}

When implementing node2vec, we set the dimension of the embedding vector as 80 for \textsc{contact}, \textsc{enron} and \textsc{radoslaw} which have less than 500 nodes. And for \textsc{fb-forum} and \textsc{lkml} with larger size, we set the dimension as 256. We grid search over \{0.5, 1, 1.5, 2\} to find the optimal values for hyper-parameters $p$ and $q$, and then use Weighted-L2~\cite{grover2016node2vec} to obtain the vector $e^{(u,v)}$ for each pair of nodes $u$ and $v$, with each element defined as
\begin{equation}
\label{eq:link feature}
e^{(u,v)}_{i} = |u_{i} - v_{i}|^{2},
\end{equation}
where $u_{i}$ and $v_{i}$ are the $i^{th}$ element of embedding vectors of nodes $u$ and $v$, respectively. For TNE, we set the dimension as 80 for \textsc{contact}, \textsc{enron} and \textsc{radoslaw} and 200 for \textsc{fb-forum} and \textsc{lkml}. The parameters of ctRBM and GTRBM are mainly about the numbers of visible units and hidden units in tRBM. The number of visible units always equals to the number of corresponding network's nodes and we set the dimension of hidden layers as 128 for smaller datasets like \textsc{contact}, \textsc{enron} and \textsc{radoslaw} and 256 for the rest. For DDNE, we set the dimension as 128 for the first three smaller datasets and 512 for the rest. When implementing our proposed model, E-LSTM-D, we choose the parameters accordingly: For the first three smaller datasets, we set $K=1$ and $L=2$ and add an additional layer to both encoder and decoder when for the rest two larger datasets. The details of the parameters are illustrated in TABLE~\ref{parameter}. Note that these parameters are chosen to get the best performance for each method, so as to make fair comparison.

\begin{table}[!t]
\newcommand{\tabincell}[2]{\begin{tabular}{@{}#1@{}}#2\end{tabular}}
\renewcommand{\arraystretch}{1.3}
\caption{The parameters of E-LSTM-D in the 5 datasets.}
\label{parameter}
\centering
\begin{tabular}{cccc}
\hline \hline
Dataset                 & \tabincell{c}{No. units in\\encoder}  & \tabincell{c}{No. units in\\stacked LSTM}  & \tabincell{c}{No. units in\\decoder}\\
\hline
\textsc{contact}        & 128              & 256~$\vert$~256          & 274         \\
\textsc{enron}          & 128              & 256~$\vert$~256          & 151         \\
\textsc{radoslaw}       & 128              & 256~$\vert$~256          & 167         \\
\textsc{fb-forum}       & 512~$\vert$~256    & 384~$\vert$~384        & 256~$\vert$~899     \\
\textsc{lkml}           & 1024~$\vert$~512   & 384~$\vert$~384        & 512~$\vert$~2210    \\
\hline \hline
\end{tabular}
\end{table}

\begin{table*}[!t]
\renewcommand{\arraystretch}{1.3}
\caption{DNLP performances on AUC, GMAUC and Error Rate for the first 20 samples and all the 80 samples.}
\centering
\label{table:performance}
\begin{tabular}{c|c|c|c|c|c|c|c|c|c|c|c}
\hline
\hline
\multirow{2}{*}{\begin{tabular}[c]{@{}c@{}}Performance \\ metric\end{tabular}} & \multirow{2}{*}{Method} & \multicolumn{2}{c|}{\textsc{contact}}  & \multicolumn{2}{c|}{\textsc{enron}} & \multicolumn{2}{c|}{\textsc{radoslaw}} & \multicolumn{2}{c|}{\textsc{fb-forum}} & \multicolumn{2}{c}{\textsc{lkml}}\\ \cline{3-12}
                               &           & 20      & 80       & 20      & 80      & 20      & 80      & 20      & 80      & 20      & 80      \\ \hline
\multirow{5}{*}{AUC}           & node2vec & 0.5212  & 0.5126  & 0.7659  & 0.6806  & 0.6103  & 0.7676  & 0.5142  & 0.5095  & 0.6348  & 0.5892  \\
          & TNE   & 0.9443  & 0.9297  & 0.8096  & 0.8314  & 0.8841  & 0.8801  & \textbf{0.9810}  & \textbf{0.9749}  & \textbf{0.9861}  & \textbf{0.9867}  \\
          & ctRBM & 0.9385  & 0.9109  & 0.8468  & 0.8295  & 0.8834  & 0.8590  & 0.8728  & 0.8349  & 0.8091  & 0.7729  \\
          & GTRBM & 0.9451  & 0.9327  & 0.8527  & 0.8491  & 0.9237  & 0.9104  & 0.9023  & 0.8749  & 0.8547  & 0.8329  \\
          & DDNE  & 0.9347  & 0.9433  & 0.7985  & 0.7638  & 0.9027  & 0.8974  & 0.9238  & 0.8729  & 0.9328  & 0.9115  \\
          & E-LSTM-D & \textbf{0.9908} & \textbf{0.9893} & \textbf{0.8931} & \textbf{0.8734} & \textbf{0.9814} & \textbf{0.9782} & 0.9670 & 0.9650 & 0.9572 & 0.9553 \\ \hline
\multirow{5}{*}{GMAUC}        & node2vec & 0.1805  & 0.1398  & 0.4069  & 0.5417  & 0.7241  & 0.7203  & 0.2744  & 0.2886  & 0.2309  & 0.2193  \\
          & TNE   & 0.9083  & 0.8958  & 0.8233  & 0.7974  & 0.8282  & 0.8251  & 0.9689  & 0.9629  & \textbf{0.9839 } & \textbf{0.9778} \\
          & ctRBM & 0.9126  & 0.8893  & 0.7207  & 0.6921  & 0.8004  & 0.7998  & 0.8926  & 0.8632  & 0.7723  & 0.7206  \\
          & GTRBM & 0.9240  & 0.9136  & 0.9148  & 0.8675  & 0.9157  & 0.8849  & 0.9329  & 0.9117  & 0.6529  & 0.6038  \\
          & DDNE  & 0.8925  & 0.8684  & 0.8724  & 0.8476  & 0.8938  & 0.8724  & 0.9126  & 0.9023  & 0.7894  & 0.7809  \\
          & E-LSTM-D & \textbf{0.9940} & \textbf{0.9902} & \textbf{0.9077} & \textbf{0.8763} & \textbf{0.9956} & \textbf{0.9938} & \textbf{0.9926} & \textbf{0.9865} & 0.8657  & 0.8511  \\ \hline
\multirow{5}{*}{Error Rate}    & node2vec & 44.7753  & 25.2278  & 23.9053  & 24.8060  & 20.7240  & 21.2489  & 40.5109  & 48.5376  & 53.2895  & 61.0274  \\
          & TNE   & 13.1410  & 7.1556  & 23.1276  & 19.9167  & 16.7078  & 16.7175  & 19.1058  & 24.4350  & 18.5702  & 18.2091  \\
          & ctRBM & 1.8976  & 1.9046  & 2.4890  & 2.7328  & 1.8920  & 2.0937  & 3.4509  & 3.6782  & 2.9903  & 3.3089  \\
          & GTRBM & 1.5843  & 1.6953  & 1.5947  & 1.8836  & 1.9079  & 2.0031  & 2.2347  & 2.4396  & 2.5351  & 2.7942  \\
          & DDNE  & 1.1780  & 1.6036  & 1.7664  & 1.9014  & 1.6316  & 1.5941  & 1.9014  & 1.8266  & 2.0134  & 2.2258  \\
          & E-LSTM-D & \textbf{0.4011} & \textbf{0.5735} & \textbf{0.9038} & \textbf{0.9880} & \textbf{0.3392} & \textbf{0.3938} & \textbf{0.5583} & \textbf{0.5777} & \textbf{0.9840} & \textbf{1.0093} \\ \hline\hline
\end{tabular}
\end{table*}

\subsection{Evaluation Metrics}
There are few metrics specifically designed for the evaluation of DNLP. Usually, those evaluation metrics used in static link prediction are also employed for DNLP. The Area Under the ROC Curve (AUC) is commonly used to measure the performance of a dynamic link predictor. AUC equals to the probability that the predictor gives a higher score to a randomly chosen existing link than a randomly chosen nonexistent one. The predictor is considered more informative if its AUC value is closer to 1. Other measurements, such as precision, Mean Average Precision (MAP), F1-score and accuracy evaluate link prediction methods from the perspective of binary classification. All of them suffer from the sparsity problem and cannot give measurements to dynamic performances. The Area Under the Precision-Recall Curve (PRAUC)~\cite{yang2015evaluating} developed from AUC is designed to deal with the sparsity of networks. However, the removed links in the near future, as a significant aspect of DNLP, are not characterized by PR curve and thus PRAUC may lose its effectiveness in this case. Junuthula et al.~\cite{junuthula2016evaluating} restricted the measurements to only part of node pairs and proposed the Geometric Mean of AUC and PRAUC (GMAUC) for the added and removed links, which can better reflect the dynamic performance. Li et al.~\cite{li2014deep} use SumD that counts the differences between the predicted network and the true one, evaluating link prediction methods in a more strict way. But the absolute difference could be misleading. For example, two dynamic link predictors both achieve SumD at 5. However, one predictor mispredicts 5 links in 10, while the other mispredicts 5 in 100. It's obvious that the latter one performs better than the former one but SumD cannot tell.

In our experiments, we choose AUC and GMAUC, and also define a new metric, \emph{Error Rate}, to evaluate our E-LSTM-D model and other baseline methods.
\begin{itemize}
\item \textbf{AUC}: If among $n$ independent comparisons, there are $n'$ times that the existing link gets a higher score than the nonexistent link and $n''$ times they get the same score, then we have
    \begin{equation}
    \text{AUC} = \frac{n'+0.5n''}{n}.
    \end{equation}
Before calculation, we randomly sample nonexistent links with the same number of existing links to ease the impact of sparsity.
\item \textbf{GMAUC}: It is a metric specifically designed for measuring the performance of DNLP. It combines PRAUC(the area under the Precision-Recall curve) and AUC by taking geometric mean of the two quantities, which is defined as
\begin{equation}
\begin{split}
\text{GMAUC} = & \bigg(\frac{{\text{PRAUC}}_{new}-\frac{L_{A}}{L_{A}+L_{R}}}{1-\frac{L_{A}}{L_{A}+L_{R}}}\\
        & ~\cdot{2({\text{AUC}}_{prev}-0.5)\bigg)}^{1/2}\\
\end{split}
\end{equation}
where $L_{A}$ and $L_{R}$ refer to the numbers of added and removed edges, respectively. $\text{PRAUC}_{new}$ is the PRAUC value calculated among the new links and $\text{AUC}_{prev}$ represents the AUC for the observed links.
\item \textbf{Error Rate}: It is defined as the ratio of the number of mispredicted links, denoted by $N_{false}$, to the total number of truly existing links, denoted by $N_{true}$, which is represented by
    \begin{equation}
    \label{eq:error rate}
    \text{Error~Rate} = \frac{N_{false}}{N_{true}}.
    \end{equation}
    Different from SumD that only counts the absolute different links in two graphs, Error Rate takes the number of truly existing links into consideration to avoid deceits.
\end{itemize}



\subsection{Experimental Results}
For each epoch, we feed 10 historical snapshots, \{$G_{t-10}$, ..., $G_{t-1}$\} to E-LSTM-D and infer $G_{t}$. And it is the same for implementing the other four DNLP approaches. For the methods that are not able to deal with time dependencies, i.e. node2vec, there are following two typical treatments: 1) only using $G_{t-1}$ to infer $G_{t}$~\cite{yao2016link}; or 2) aggregating previous 10 snapshots into a single network and then do link prediction~\cite{nguyen2018continuous,li2018deep}. We choose the former one when implementing node2vec, because the relatively long sequence of historical snapshots here may carry some disturbing information that node2vec cannot handle, leading to even poor performance.

We compare our E-LSTM-D model with the five baseline methods on the performance metrics AUC, GMAUC and Error Rate. Since the patterns of network evolution may change with time, the model trained by the history data may not capture the pattern in the remote future. To investigate both short-term and long-term prediction performance, we report the average values of the three performance metrics for both the first 20 test samples and all the 80 samples. The results are presented in TABLE~\ref{table:performance}, where we can see that, generally, the E-LSTM-D model outperforms all the baseline methods in almost all the cases, no matter the network is large or small, dense or sparse, for both short-term and long-term prediction. In particular, for the metrics of AUC and GMAUC, the poor performances obtained by node2vec indicate that the methods, designed for static networks, are indeed not suitable for DNLP. On the contrary, E-LSTM-D and other DNLP baselines can get much better performances, due to their dynamic nature. 

\begin{figure*}[!t]
\centering
\includegraphics[width=1\linewidth]{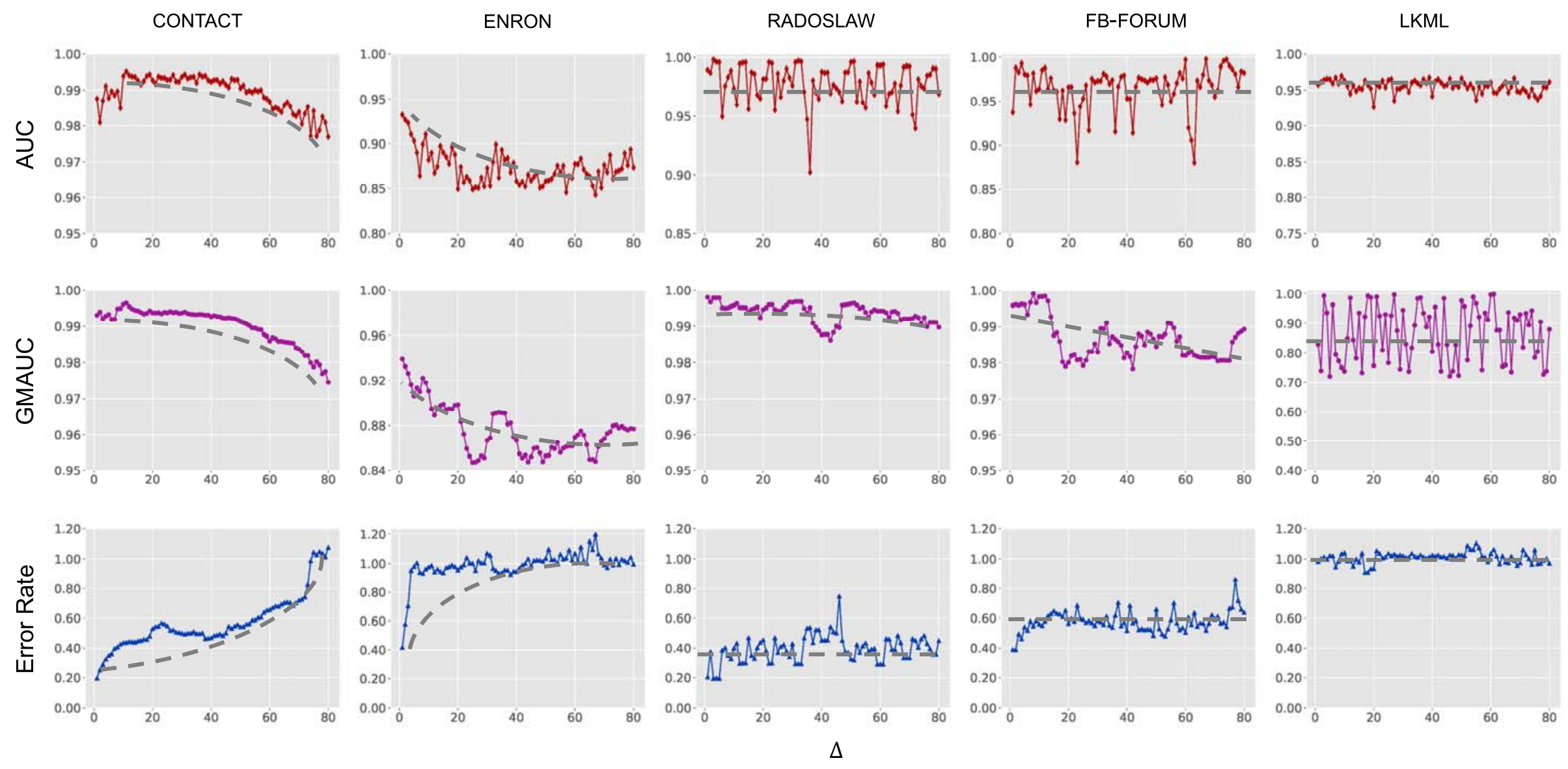}
\caption{DNLP performance on AUC, GMAUC and Error Rate, obtained by our E-LSTM-D model, as functions of $\Delta$ for the five datasets. The dash lines represent the changing tendencies.}
\label{fig:tendency}
\end{figure*}


\begin{table*}[!t]
\newcommand{\tabincell}[2]{\begin{tabular}{@{}#1@{}}#2\end{tabular}}
\renewcommand{\arraystretch}{1.3}
\caption{Error Rate of the top 10\% important links, in terms of DC and EBC, for the first 20 test samples and all the 80 samples.}
\centering
\label{table:important edges}
\begin{tabular}{c|c|c|c|c|c|c|c|c|c|c|c}
\hline\hline
\multirow{2}{*}{\begin{tabular}[c]{@{}c@{}}Metric \\ for link importance\end{tabular}} & \multirow{2}{*}{Method} & \multicolumn{2}{c|}{\textsc{contact}} & \multicolumn{2}{c|}{\textsc{enron}} & \multicolumn{2}{c|}{\textsc{radoslaw}} & \multicolumn{2}{c|}{\textsc{fb-forum}} & \multicolumn{2}{c}{\textsc{lkml}}\\ \cline{3-12}
                        &           & 20      & 80      & 20      & 80      & 20      & 80      & 20      & 80      & 20      & 80             \\ \hline
\multirow{5}{*}{\tabincell{c}{Degree\\centrality}}     & node2vec & 0.6279  & 0.6297  & 0.4900  & 0.4524  & 0.4735  & 0.5203  & 0.3873  & 0.3454  & 0.5034  & 0.5289  \\
          & TNE   & 0.9622  & 0.9551  & 0.3446  & 0.3315  & 0.5068  & 0.4413  & 0.0595  & 0.0558  & 0.6390  & 0.6288  \\
          & ctRBM & 0.2739  & 0.3307  & 0.4193 & 0.4410  & 0.3028  & 0.3097  & 0.1095  & 0.1137  & 0.3291  & 0.3341  \\
          & GTRBM & 0.2209  & 0.2390  & 0.4098  & 0.4322  & 0.2109  & 0.2198  & 0.1127  & 0.1239  & 0.2973  & 0.3030  \\
          & DDNE  & 0.1293  & 0.1359  & 0.2270  & 0.2133  & 0.0803  & 0.1249  & 0.1190  & 0.1088  & \textbf{0.1653} & \textbf{0.1821} \\
          & E-LSTM-D & \textbf{0.0484 } & \textbf{0.1109} & \textbf{0.2182} & \textbf{0.2096} & \textbf{0.0516} & \textbf{0.0761} & \textbf{0.0160} & \textbf{0.0222} & 0.1863  & 0.1992\\ \hline
\multirow{5}{*}{\tabincell{c}{Edge betweenness\\centrality}}    & node2vec & 0.6747 & 0.6509  & 0.4607  & 0.5953  & 0.4657  & 0.4397  & 0.6517  & 0.6799  & 0.8729  & 0.8698  \\
          & TNE   & 0.9998 & 0.9987  & 0.9598  & 0.9590  & 1.0000  & 1.0000  & 1.0000  & 0.9986  & 1.0000  & 0.9992  \\
          & ctRBM & 0.5396  & 0.5619  & 0.6512  & 0.7381  & 0.2165  & 0.2291  & 0.4432  & 0.4508  & 0.7279  & 0.7503  \\
          & GTRBM & 0.4418  & 0.4573  & 0.6906  & 0.7420  & 0.2399  & 0.2511  & 0.4507  & 0.4529  & 0.6370  & 0.6524  \\
          & DDNE  & 0.2713  & 0.2849  & \textbf{0.4988} & \textbf{0.5471} & 0.2083  & 0.2508  & 0.2697  & 0.3014  & 0.6435  & 0.6614  \\
          & E-LSTM-D & \textbf{0.2004} & \textbf{0.2547} & 0.5067  & 0.6157  & \textbf{0.1617} & \textbf{0.2159} & \textbf{0.2643} & \textbf{0.2825} & \textbf{0.5820} & \textbf{0.6126} \\ \hline\hline
\end{tabular}
\end{table*}

\begin{figure}[!t]
\centering
\includegraphics[width=1\linewidth]{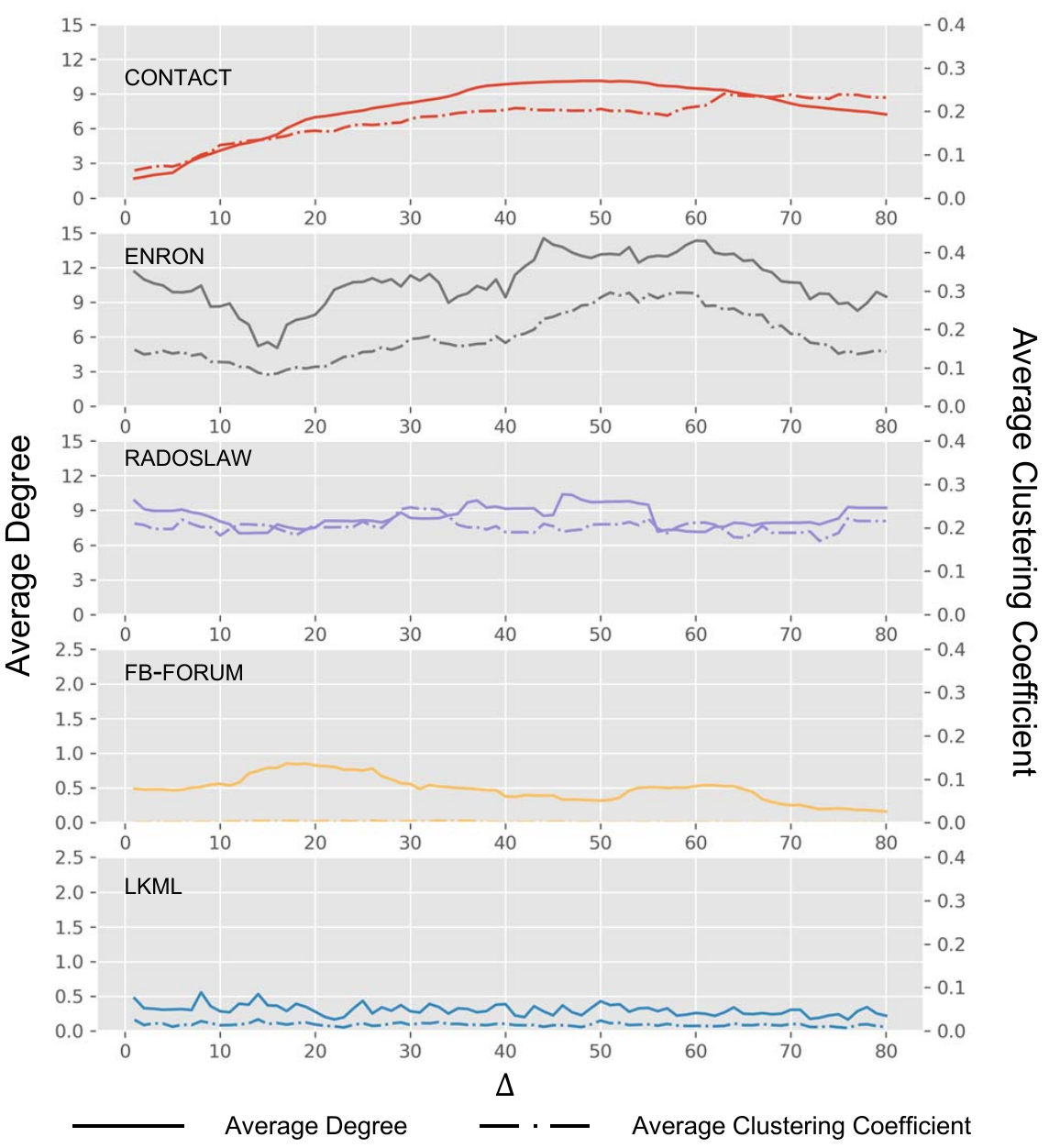}
\caption{The network structural properties, average degree and average clustering coefficient, as functions of $\Delta$ for the five datasets.}
\label{fig:network_structure}
\end{figure}

Moreover, for each predicted snapshot, we also compare the predicted links with truly existing ones to obtain the Error Rate. We find that node2vec can easily predict much more links than the truly existing ones, leading to relatively large Error Rates. We argue that it might blame to the classification process that the pre-trained linear regression model is not suitable for the classification of embedding vectors. As presented in TABLE~\ref{table:performance}, the results again demonstrate the best performance of our E-LSTM-D model on DNLP. TNE performaces poorly on Error Rate, because it does not specially fit the distribution of the network as the other deep learning based methods do. The dramatic difference of the Error Rate between E-LSTM-D and TNE indicates that this metric is a good addition to AUC to comprehensively measure the performance of DNLP. Other deep learning based methods, like ctRBM and DDNE, have similar performances while they could not compete with E-LSTM-D in most cases. It is worth noticing that the TNE outperforms the others on \textsc{lkml} from the perspective of traditional AUC and GMAUC, which shows its robustness to the scale of networks on these metrics, however, it has much larger Error Rate compared with the other DNLP methods.


For the 80 test samples with $G_{240+\Delta}$ as the output, where $\Delta$ varies from 1 to 80, we draw the DNLP performances on the three metrics, obtained by E-LSTM-D, as functions of $\Delta$ for the five datasets to see how long it can predict network evolution with satisfying performance. The results are shown in Fig.~\ref{fig:tendency} for E-LSTM-D, where we can see that, generally, AUC and GMAUC decrease, while Error Rate increases, as $t$ increases, indicating that long-term prediction on structure is indeed relatively difficult for most dynamic networks. Interestingly, for \textsc{radoslaw}, \textsc{fb-forum} and \textsc{lkml}, the prediction performances are relatively stable, which might be because their network structures evolve periodically, making the collection of snapshots easy to predict, especially when LSTM is integrated in our deep learning framework. To further illustrate this, we investigate the changing trends of the most common structural properties, i.e., average degree and average clustering coefficient, of the five networks as $\Delta$ increases. The results are shown in Fig.~\ref{fig:network_structure}, where we can see that these two properties change dramatically for \textsc{contact} and \textsc{enron}, while they are relatively stable for \textsc{radoslaw}, \textsc{fb-forum} and \textsc{lkml}. These results explain why we can make better long-term prediction on the last two dynamic networks.



As described above, although some methods have excellent performances on AUC, they might mispredict many links. In most real-world scenarios, however, we may only focus on the most important links. Therefore, we further evaluate our model on part of the links that are of particular significance in the network. Here, we use two metrics, degree centrality and edge betweenness centrality, to measure the importance of each link. DC is originally used to measure the importance of nodes according to the amount of neighbors. To measure the importance of a link, we use the sum of degree centralities of the two terminal nodes (source and target). We then calculate the Error Rate when predicting the top 10\% important links. The results are presented in TABLE~\ref{table:important edges}, which demonstrate again the outstanding performance of our E-LSTM-D model in predicting important links. It also shows that the E-LSTM-D model is more capable of learning networks' features, i.e. degree distribution and edge betweenness, which could account for the effectiveness in a way. Moreover, comparing TABLE.~\ref{table:performance} and TABLE.~\ref{table:important edges}, we find that Error Rates on the top 10\% important links are much smaller than those on all the links in the five networks by adopting any method. This indicates that, actually, those more important links are also more easily to be predicted.


\subsection{Beyond Link Prediction}
\begin{figure}[!t]
\centerline{\includegraphics[width=1.\linewidth]{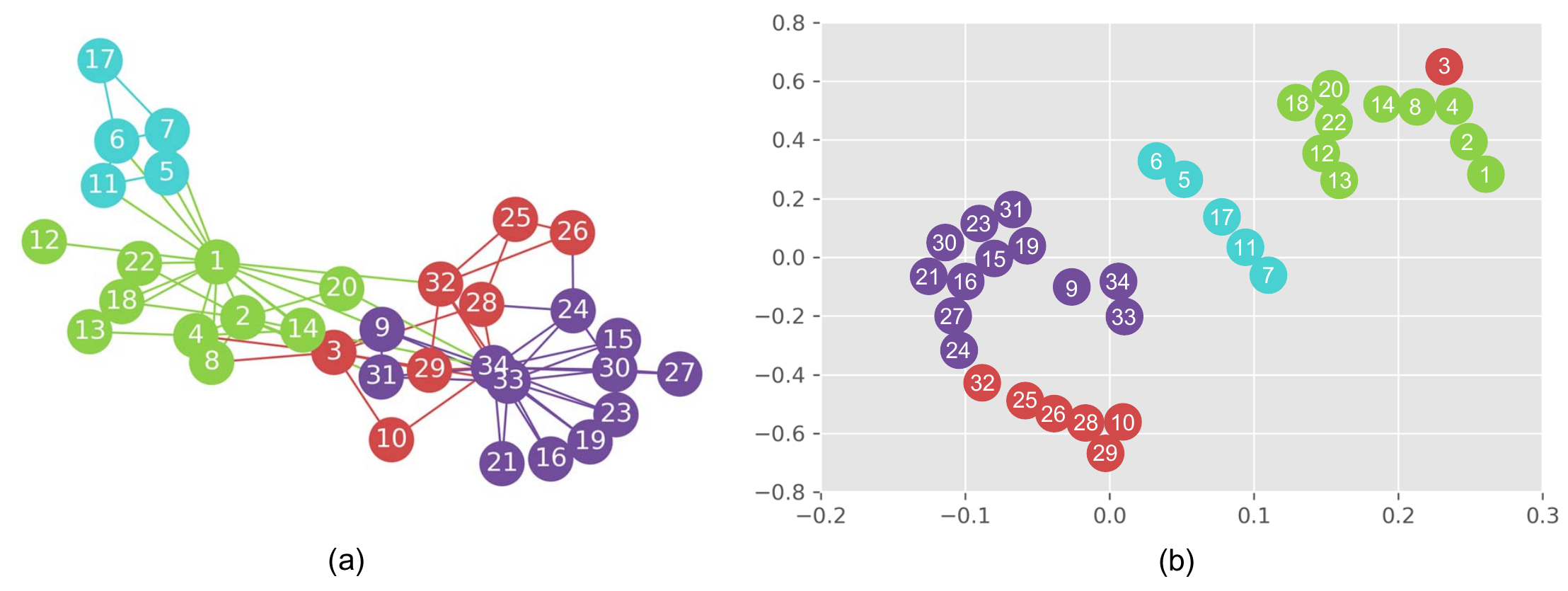}}
\caption{(a) The structure of karate club network. (b) The t-SNE visualization of the embedding features obtained by our E-LSTM-D model.}
\label{fig:visualization}
\end{figure}

Our E-LSTM-D model learns low dimensional representation for each node in the process of link prediction. These vectors, like those generated by other network embedding methods, contains local or global structural information that can be used in other tasks such as node classification etc. To illustrate this, we conduct experiment on karate club dataset, with the network structure shown in Fig.~\ref{fig:visualization} (a). We first obtain $G_{t-1}$ by randomly removing 10 links form the original network and then use it to predict the original network $G_{t}$. After training, we use the output of the stacked LSTM as the input to the visualization method t-SNE~\cite{maaten2008visualizing}. Besides obtaining the excellent performance on link prediction, we also visualize the embedding vectors, as shown in Fig.~\ref{fig:visualization} (b), where we can see that the nodes of the same class are close to each other while those of different classes are relatively far away. This indicates that the embedding vectors obtained by our E-LSTM-D model on link prediction can also be used to effectively solve the node classification problem, validating the outstanding transferability of the model.

\subsection{Parameter Sensitivity}
The performance of our E-LSTM-D model is mainly determined by three parts: the structure of model, the length of historical snapshots $N$, and the penalty coefficient $\beta$. In the following, we will investigate their influences on the model performance.

\begin{figure}[!t]
\centering
\includegraphics[width=1\linewidth]{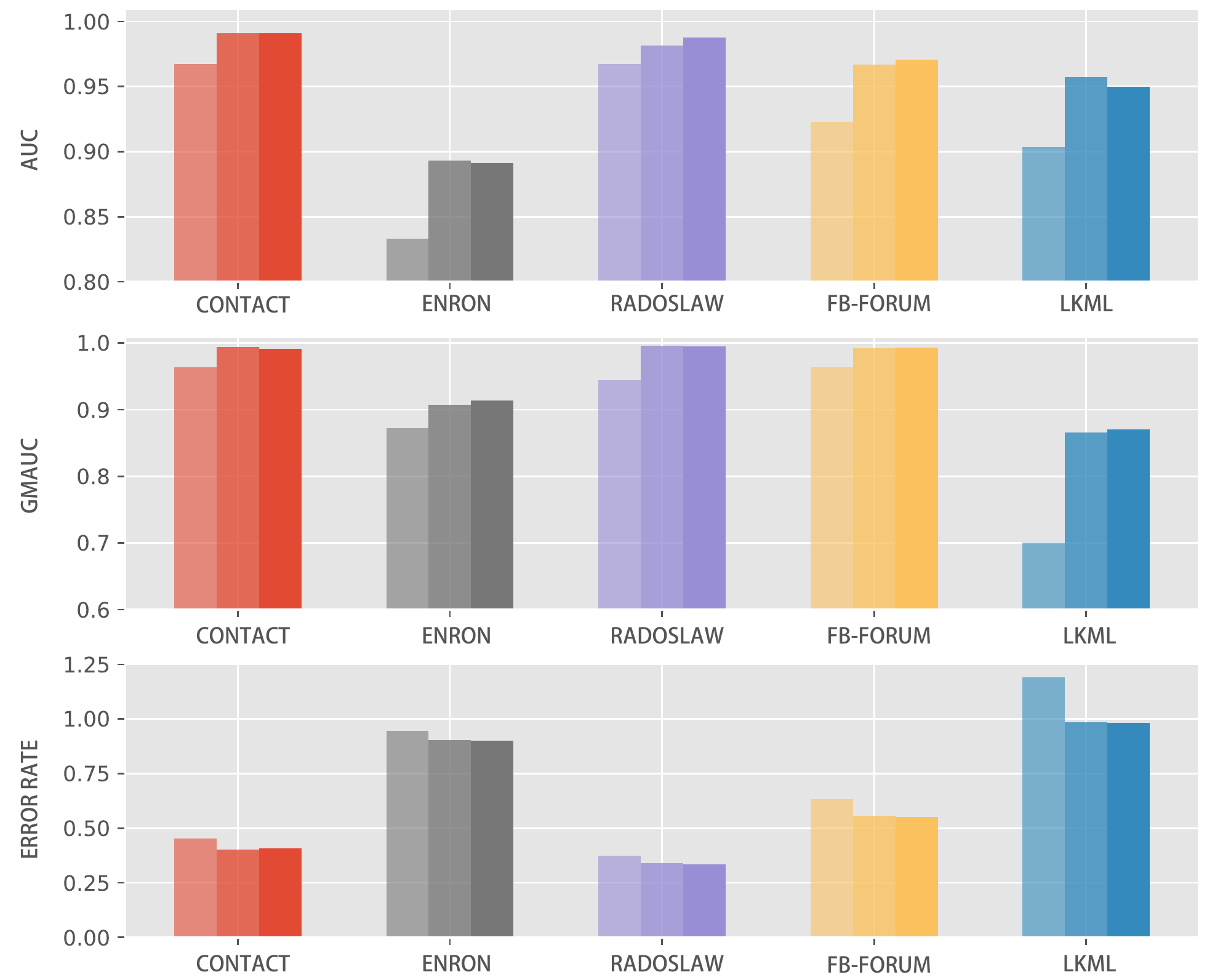}
\caption{Performance on the 5 datasets with different number of units of the first encoder layer. For each dataset, the number of the units of the first encoder layer increases at the step of 64 from left to right.}
\label{fig:parameter_units}
\end{figure}

\begin{figure*}[!t]
\centering
\includegraphics[width=1\linewidth]{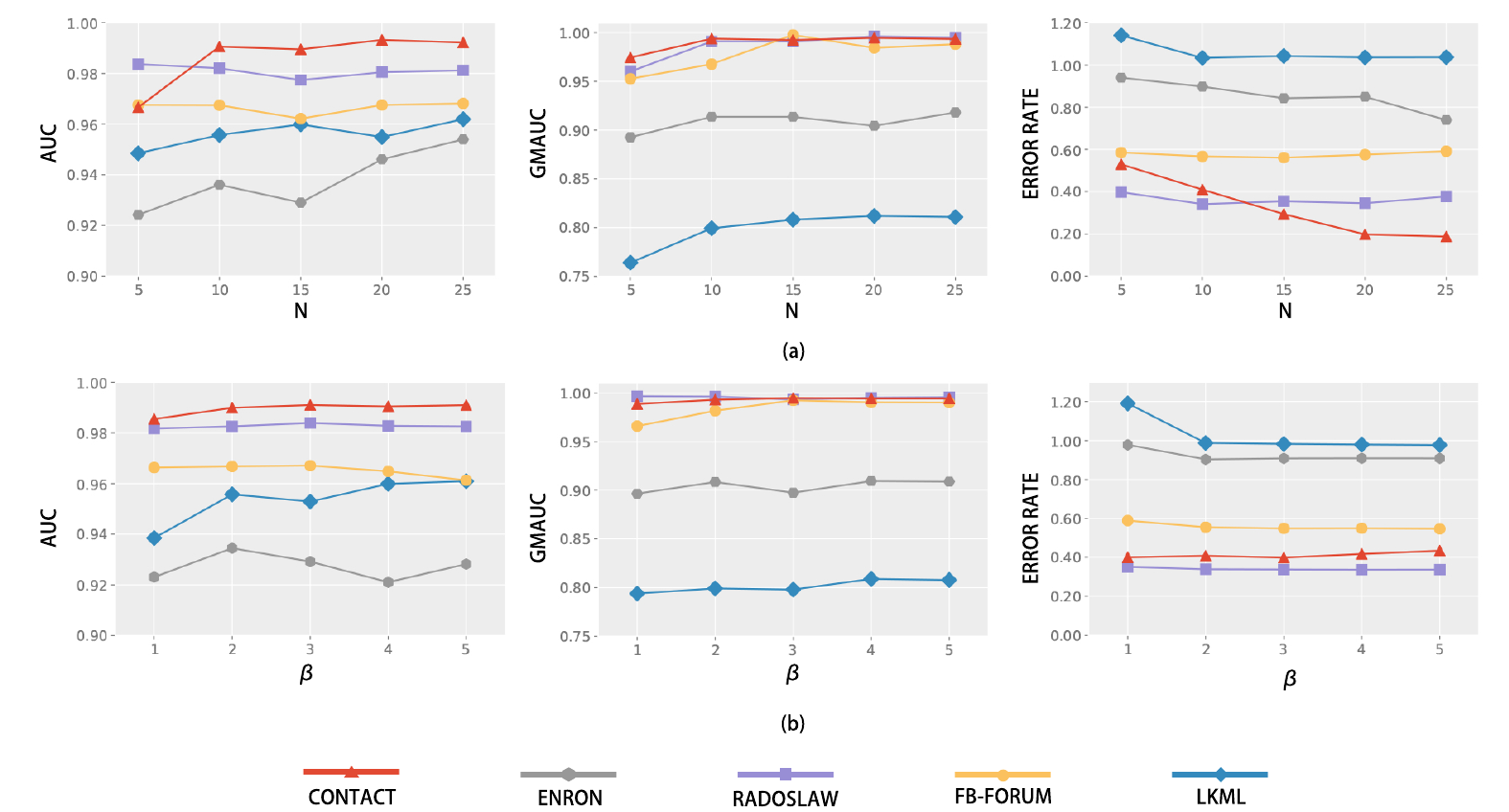}
\caption{Parameter sensitivity analysis of our E-LSTM-D model on five datasets. (a) The performances on AUC, GMAUC and Error Rates as functions of historical snapshot length $N$. (b) The performances as functions of penalty coefficient $\beta$.}
\label{fig:parameter}
\end{figure*}

\subsubsection{Influence of the model's structure}
The results shown in TABLE~\ref{table:performance} are obtained by the models with selected structures. The numbers of units in each layer and the number of layers are set with concerns on both computation complexity and models' performance. We test the model with different number of units and encoder layers to prove the validity of the structures above. Fig.~\ref{fig:parameter_units} shows that the performance will slightly drop with the reduction of the number of units in the first encoder layer. And further increasing the complexity has little contribution to the performance and may even lead to worse results. TABLE~\ref{tab:parameter_layer} reports the difference of the performances between the model with an additional encoder layer which shares the same structure of the previous layer and the original model. The results show that there seems no significant improvements on AUC and GMAUC with an additional layer. But it could actually lower Error Rates with the increasing of the model's complexity. Overall, the general structure of E-LSTM-D can achieve state-of-art performance in most cases.

\begin{table}[!t]
\centering
\renewcommand{\arraystretch}{1.6}
\caption{Difference of the performance with different number of encoder layers}
\label{tab:parameter_layer}
\begin{tabular}{c|c|c|c}
\hline \hline
                  & AUC      & GMAUC      & ERROR RATE \\ \hline
\textsc{contact}  & 0.0038   & 0.0024     & -0.1037    \\
\textsc{enron}    & 0.0119   & 0.0206     & -0.0397    \\
\textsc{radoslaw} & 0.0035   & -0.0054    & -0.0920    \\
\textsc{fb-forum} & 0.0029   & 0.0011     & -0.1033    \\
\textsc{lkml}     & -0.0079  & 0.0108     & -0.1375    \\
\hline \hline
\end{tabular}
\end{table}

\subsubsection{Influence of historical snapshot length}
Usually, longer length of historical snapshots contains more information and thus may improve link prediction performance. On the other hand, snapshots from long ago, however, might have little influence on the current snapshot, while more historical snapshots will increase the computational complexity. Therefore, it is necessary to find a proper length to balance efficiency and performance. We thus vary the length of historical snapshots from 5 to 25 with a regular interval 5. The results are shown in Fig.~\ref{fig:parameter} (a), which tell that more historical snapshots can indeed improve the performance of our model, i.e., leading to larger AUC and GMAUC while smaller Error Rate. Moreover, it seems that AUC and GMAUC increase most when $N$ changes from 1 to 10, while Error Rate decreases most when $N$ changes from 1 to 20. Thereafter, for most dynamic networks, these metrics keep almost the same as $N$ further increases. This phenomenon suggests us to choose $N=10$ in the previous experiments.

\subsubsection{Influence of the penalty coefficient}
The penalty coefficient $\beta$ is applied in the objective to avoid overfitting and accelerate convergence. When $\beta=1$, the objective simply equals to $L_2$ distance. In reality, $\beta$ is usually larger than 1 to help the model focus more on existing links in the training process. As shown in Fig.~\ref{fig:parameter} (b), we can see that the performance is relatively stable as $\beta$ varies. However, for some datasets, the increasing of penalty coefficient could actually lead to slightly larger GMAUC but smaller Error Rate, while it has little effect on AUC. As $\beta$ further increases, both GMAUC and Error Rate keep relatively stable. These suggest us to choose a relatively small $\beta$, i.e., $\beta\in(1,2]$ in the experiments, varying for different datasets to obtain the optimal results.

\section{Conclusion\label{Con}}
In this paper, we propose a new deep learning model, namely E-LSTM-D, for DNLP. Specifically, to predict future links, we design an end-to-end model integrating a stacked LSTM into the architecture of encoder-decoder, which can make fully use of historical information. The proposed model learns not only the low dimensional representations and non-linearity but also the time dependencies between successive network snapshots, as a result, it can better capture the patterns of network evolution. To cope with the problem of sparsity, we impose more penalty to exis links in the objective, which can also help to preserve local structure and accelerate convergence. Empirically, we conduct extensive experiments to compare our model with traditional link prediction methods on a variety of datasets. The results demonstrate that our model outperforms the others and achieve the state-of-the-art performance. Moreover, we show that the latent features generated by our model in link prediction can be used to well characterize the global and local structure of the nodes in a network and thus may also benefit other tasks, such as node classification.

Our future research will focus on predicting the evolution of layered dynamic networks. Besides, we will make efforts to reduce the computational complexity of our E-LSTM-D model to make it suitable for large-scale network. Also, we will study the transferability of our model on various tasks by conducting more comprehensive experiments.

\bibliographystyle{IEEEtran}
\bibliography{ref}

\begin{thebibliography}{10}
\providecommand{\url}[1]{#1}
\csname url@samestyle\endcsname
\providecommand{\newblock}{\relax}
\providecommand{\bibinfo}[2]{#2}
\providecommand{\BIBentrySTDinterwordspacing}{\spaceskip=0pt\relax}
\providecommand{\BIBentryALTinterwordstretchfactor}{4}
\providecommand{\BIBentryALTinterwordspacing}{\spaceskip=\fontdimen2\font plus
\BIBentryALTinterwordstretchfactor\fontdimen3\font minus
  \fontdimen4\font\relax}
\providecommand{\BIBforeignlanguage}[2]{{%
\expandafter\ifx\csname l@#1\endcsname\relax
\typeout{** WARNING: IEEEtran.bst: No hyphenation pattern has been}%
\typeout{** loaded for the language `#1'. Using the pattern for}%
\typeout{** the default language instead.}%
\else
\language=\csname l@#1\endcsname
\fi
#2}}
\providecommand{\BIBdecl}{\relax}
\BIBdecl

\bibitem{ediger2010massive}
D.~Ediger, K.~Jiang, J.~Riedy, D.~A. Bader, and C.~Corley, ``Massive social
  network analysis: Mining twitter for social good,'' in \emph{Parallel
  Processing (ICPP), 2010 39th International Conference on}.\hskip 1em plus
  0.5em minus 0.4em\relax IEEE, 2010, pp. 583--593.

\bibitem{fu2017pinning}
C.~Fu, J.~Wang, Y.~Xiang, Z.~Wu, L.~Yu, and Q.~Xuan, ``Pinning control of
  clustered complex networks with different size,'' \emph{Physica A:
  Statistical Mechanics and its Applications}, vol. 479, pp. 184--192, 2017.

\bibitem{wang2017investigating}
L.~Wang and J.~Orchard, ``Investigating the evolution of a neuroplasticity
  network for learning,'' \emph{IEEE Transactions on Systems, Man, and
  Cybernetics: Systems}, 2018, \mbox{doi}:\url{10.1109/TSMC.2017.2755066}.

\bibitem{GAO2012391}
J.~Gao, Y.~Xiao, J.~Liu, W.~Liang, and C.~P. Chen, ``A survey of
  communication/networking in smart grids,'' \emph{Future Generation Computer
  Systems}, vol.~28, no.~2, pp. 391 -- 404, 2012.

\bibitem{kazemilari2015correlation}
M.~Kazemilari and M.~A. Djauhari, ``Correlation network analysis for
  multi-dimensional data in stocks market,'' \emph{Physica A: Statistical
  Mechanics and its Applications}, vol. 429, pp. 62--75, 2015.

\bibitem{sun2017complex}
J.~Sun, Y.~Yang, N.~N. Xiong, L.~Dai, X.~Peng, and J.~Luo, ``Complex network
  construction of multivariate time series using information geometry,''
  \emph{IEEE Transactions on Systems, Man, and Cybernetics: Systems}, vol.~49,
  no.~1, pp. 107--122, Jan 2019.

\bibitem{liu2018optimizing}
H.~Liu, X.~Xu, J.-A. Lu, G.~Chen, and Z.~Zeng, ``Optimizing pinning control of
  complex dynamical networks based on spectral properties of grounded laplacian
  matrices,'' \emph{IEEE Transactions on Systems, Man, and Cybernetics:
  Systems}, 2018, \mbox{doi}:\url{10.1109/TSMC.2018.2882620}.

\bibitem{ibrahim2015link}
N.~M.~A. Ibrahim and L.~Chen, ``Link prediction in dynamic social networks by
  integrating different types of information,'' \emph{Applied Intelligence},
  vol.~42, no.~4, pp. 738--750, 2015.

\bibitem{xuan2015temporal}
Q.~Xuan, H.~Fang, C.~Fu, and V.~Filkov, ``Temporal motifs reveal collaboration
  patterns in online task-oriented networks,'' \emph{Physical Review E},
  vol.~91, no.~5, p. 052813, 2015.

\bibitem{xuan2018social}
Q.~Xuan, Z.-Y. Zhang, C.~Fu, H.-X. Hu, and V.~Filkov, ``Social synchrony on
  complex networks,'' \emph{IEEE transactions on cybernetics}, vol.~48, no.~5,
  pp. 1420--1431, 2018.

\bibitem{xuan2018modern}
Q.~Xuan, M.~Zhou, Z.-Y. Zhang, C.~Fu, Y.~Xiang, Z.~Wu, and V.~Filkov, ``Modern
  food foraging patterns: Geography and cuisine choices of restaurant patrons
  on yelp,'' \emph{IEEE Transactions on Computational Social Systems}, vol.~5,
  no.~2, pp. 508--517, 2018.

\bibitem{fu2018link}
C.~Fu, M.~Zhao, L.~Fan, X.~Chen, J.~Chen, Z.~Wu, Y.~Xia, and Q.~Xuan, ``Link
  weight prediction using supervised learning methods and its application to
  yelp layered network,'' \emph{IEEE Transactions on Knowledge and Data
  Engineering}, 2018.

\bibitem{lentz2016disease}
H.~H. Lentz, A.~Koher, P.~H{\"o}vel, J.~Gethmann, C.~Sauter-Louis, T.~Selhorst,
  and F.~J. Conraths, ``Disease spread through animal movements: a static and
  temporal network analysis of pig trade in germany,'' \emph{PloS one},
  vol.~11, no.~5, p. e0155196, 2016.

\bibitem{theocharidis2009network}
A.~Theocharidis, S.~Van~Dongen, A.~J. Enright, and T.~C. Freeman, ``Network
  visualization and analysis of gene expression data using biolayout express
  3d,'' \emph{Nature protocols}, vol.~4, no.~10, p. 1535, 2009.

\bibitem{newman2001clustering}
M.~E. Newman, ``Clustering and preferential attachment in growing networks,''
  \emph{Physical review E}, vol.~64, no.~2, p. 025102, 2001.

\bibitem{zhou2009predicting}
T.~Zhou, L.~L{\"u}, and Y.-C. Zhang, ``Predicting missing links via local
  information,'' \emph{The European Physical Journal B}, vol.~71, no.~4, pp.
  623--630, 2009.

\bibitem{lu2011link}
L.~L{\"u} and T.~Zhou, ``Link prediction in complex networks: A survey,''
  \emph{Physica A: statistical mechanics and its applications}, vol. 390,
  no.~6, pp. 1150--1170, 2011.

\bibitem{yao2016link}
L.~Yao, L.~Wang, L.~Pan, and K.~Yao, ``Link prediction based on
  common-neighbors for dynamic social network,'' \emph{Procedia Computer
  Science}, vol.~83, pp. 82--89, 2016.

\bibitem{zhang2017efficient}
Z.~Zhang, J.~Wen, L.~Sun, Q.~Deng, S.~Su, and P.~Yao, ``Efficient incremental
  dynamic link prediction algorithms in social network,'' \emph{Knowledge-Based
  Systems}, vol. 132, pp. 226--235, 2017.

\bibitem{perozzi2014deepwalk}
B.~Perozzi, R.~Al-Rfou, and S.~Skiena, ``Deepwalk: Online learning of social
  representations,'' in \emph{Proceedings of the 20th ACM SIGKDD international
  conference on Knowledge discovery and data mining}.\hskip 1em plus 0.5em
  minus 0.4em\relax ACM, 2014, pp. 701--710.

\bibitem{grover2016node2vec}
A.~Grover and J.~Leskovec, ``node2vec: Scalable feature learning for
  networks,'' in \emph{Proceedings of the 22nd ACM SIGKDD international
  conference on Knowledge discovery and data mining}.\hskip 1em plus 0.5em
  minus 0.4em\relax ACM, 2016, pp. 855--864.

\bibitem{han2018deep}
Z.~Han, Z.~Liu, C.-M. Vong, Y.-S. Liu, S.~Bu, J.~Han, and C.~P. Chen, ``Deep
  spatiality: Unsupervised learning of spatially-enhanced global and local 3d
  features by deep neural network with coupled softmax,'' \emph{IEEE
  Transactions on Image Processing}, vol.~27, no.~6, pp. 3049--3063, 2018.

\bibitem{xuan2018automatic}
Q.~Xuan, B.~Fang, Y.~Liu, J.~Wang, J.~Zhang, Y.~Zheng, and G.~Bao, ``Automatic
  pearl classification machine based on a multistream convolutional neural
  network,'' \emph{IEEE Transactions on Industrial Electronics}, vol.~65,
  no.~8, pp. 6538--6547, 2018.

\bibitem{xuan2018evolving}
Q.~Xuan, H.~Xiao, C.~Fu, and Y.~Liu, ``Evolving convolutional neural network
  and its application in fine-grained visual categorization,'' \emph{IEEE
  Access}, 2018.

\bibitem{wang2016structural}
D.~Wang, P.~Cui, and W.~Zhu, ``Structural deep network embedding,'' in
  \emph{Proceedings of the 22nd ACM SIGKDD international conference on
  Knowledge discovery and data mining}.\hskip 1em plus 0.5em minus 0.4em\relax
  ACM, 2016, pp. 1225--1234.

\bibitem{kipf2016semi}
T.~N. Kipf and M.~Welling, ``Semi-supervised classification with graph
  convolutional networks,'' \emph{arXiv preprint arXiv:1609.02907}, 2016.

\bibitem{ahmed2016sampling}
N.~M. Ahmed, L.~Chen, Y.~Wang, B.~Li, Y.~Li, and W.~Liu, ``Sampling-based
  algorithm for link prediction in temporal networks,'' \emph{Information
  Sciences}, vol. 374, pp. 1--14, 2016.

\bibitem{ahmed2016efficient}
N.~M. Ahmed and L.~Chen, ``An efficient algorithm for link prediction in
  temporal uncertain social networks,'' \emph{Information Sciences}, vol. 331,
  pp. 120--136, 2016.

\bibitem{li2014deep}
X.~Li, N.~Du, H.~Li, K.~Li, J.~Gao, and A.~Zhang, ``A deep learning approach to
  link prediction in dynamic networks,'' in \emph{Proceedings of the 2014 SIAM
  International Conference on Data Mining}.\hskip 1em plus 0.5em minus
  0.4em\relax SIAM, 2014, pp. 289--297.

\bibitem{zhou2018dynamic}
L.~{Zhou}, Y.~{Yang}, X.~{Ren}, F.~{Wu}, and Y.~{Zhuang}, ``{Dynamic Network
  Embedding by Modelling Triadic Closure Process},'' in \emph{AAAI}, 2018.

\bibitem{li2018deep}
T.~Li, J.~Zhang, S.~Y. Philip, Y.~Zhang, and Y.~Yan, ``Deep dynamic network
  embedding for link prediction,'' \emph{IEEE Access}, 2018.

\bibitem{chung2014empirical}
J.~Chung, C.~Gulcehre, K.~Cho, and Y.~Bengio, ``Empirical evaluation of gated
  recurrent neural networks on sequence modeling,'' \emph{arXiv preprint
  arXiv:1412.3555}, 2014.

\bibitem{gers1999learning}
F.~A. Gers, J.~Schmidhuber, and F.~Cummins, ``Learning to forget: Continual
  prediction with lstm,'' 1999.

\bibitem{lipton2015critical}
Z.~C. Lipton, J.~Berkowitz, and C.~Elkan, ``A critical review of recurrent
  neural networks for sequence learning,'' \emph{arXiv preprint
  arXiv:1506.00019}, 2015.

\bibitem{han2019seqviews2seqlabels}
Z.~Han, M.~Shang, Z.~Liu, C.-M. Vong, Y.-S. Liu, M.~Zwicker, J.~Han, and C.~P.
  Chen, ``Seqviews2seqlabels: Learning 3d global features via aggregating
  sequential views by rnn with attention,'' \emph{IEEE Transactions on Image
  Processing}, vol.~28, no.~2, pp. 658--672, 2019.

\bibitem{konect:2017:contact}
\BIBentryALTinterwordspacing
``Haggle network dataset -- {KONECT},'' Apr. 2017. [Online]. Available:
  \url{http://konect.uni-koblenz.de/networks/contact}
\BIBentrySTDinterwordspacing

\bibitem{nr}
\BIBentryALTinterwordspacing
R.~A. Rossi and N.~K. Ahmed, ``The network data repository with interactive
  graph analytics and visualization,'' in \emph{Proceedings of the Twenty-Ninth
  AAAI Conference on Artificial Intelligence}, 2015. [Online]. Available:
  \url{http://networkrepository.com}
\BIBentrySTDinterwordspacing

\bibitem{konect:radoslaw}
R.~Michalski, S.~Palus, and P.~Kazienko, ``Matching organizational structure
  and social network extracted from email communication,'' in \emph{Lecture
  Notes in Business Information Processing}, vol.~87.\hskip 1em plus 0.5em
  minus 0.4em\relax Springer Berlin Heidelberg, 2011, pp. 197--206.

\bibitem{konect:2017:facebook-wosn-wall}
\BIBentryALTinterwordspacing
``Facebook wall posts network dataset -- {KONECT},'' Apr. 2017. [Online].
  Available: \url{http://konect.uni-koblenz.de/networks/facebook-wosn-wall}
\BIBentrySTDinterwordspacing

\bibitem{konect:2017:lkml-reply}
\BIBentryALTinterwordspacing
``Linux kernel mailing list replies network dataset -- {KONECT},'' Apr. 2017.
  [Online]. Available: \url{http://konect.uni-koblenz.de/networks/lkml-reply}
\BIBentrySTDinterwordspacing

\bibitem{zhu2016scalable}
L.~Zhu, D.~Guo, J.~Yin, G.~Ver~Steeg, and A.~Galstyan, ``Scalable temporal
  latent space inference for link prediction in dynamic social networks,''
  \emph{IEEE Transactions on Knowledge and Data Engineering}, vol.~28, no.~10,
  pp. 2765--2777, 2016.

\bibitem{li2018restricted}
T.~Li, B.~Wang, Y.~Jiang, Y.~Zhang, and Y.~Yan, ``Restricted boltzmann
  machine-based approaches for link prediction in dynamic networks,''
  \emph{IEEE Access}, 2018.

\bibitem{yang2015evaluating}
Y.~Yang, R.~N. Lichtenwalter, and N.~V. Chawla, ``Evaluating link prediction
  methods,'' \emph{Knowledge and Information Systems}, vol.~45, no.~3, pp.
  751--782, 2015.

\bibitem{junuthula2016evaluating}
R.~R. Junuthula, K.~S. Xu, and V.~K. Devabhaktuni, ``Evaluating link prediction
  accuracy in dynamic networks with added and removed edges,'' in \emph{Big
  Data and Cloud Computing (BDCloud), Social Computing and Networking
  (SocialCom), Sustainable Computing and Communications
  (SustainCom)(BDCloud-SocialCom-SustainCom), 2016 IEEE International
  Conferences on}.\hskip 1em plus 0.5em minus 0.4em\relax IEEE, 2016, pp.
  377--384.

\bibitem{nguyen2018continuous}
G.~H. Nguyen, J.~B. Lee, R.~A. Rossi, N.~K. Ahmed, E.~Koh, and S.~Kim,
  ``Continuous-time dynamic network embeddings,'' in \emph{3rd International
  Workshop on Learning Representations for Big Networks (WWW BigNet)}, 2018.

\bibitem{maaten2008visualizing}
L.~Maaten and G.~Hinton, ``Visualizing data using t-sne,'' \emph{Journal of
  machine learning research}, vol.~9, no. Nov, pp. 2579--2605, 2008.

\end{thebibliography}

\end{document}